\documentclass[sigconf]{acmart}%
\AtBeginDocument{%
  \providecommand\BibTeX{{%
    \normalfont B\kern-0.5em{\scshape i\kern-0.25em b}\kern-0.8em\TeX}}}

\copyrightyear{2022}
\acmYear{2022}
\setcopyright{acmcopyright}\acmConference[UIST '22]{The 35th Annual ACM Symposium on User Interface Software and Technology}{October 29-November 2, 2022}{Bend, OR, USA}
\acmBooktitle{The 35th Annual ACM Symposium on User Interface Software and Technology (UIST '22), October 29-November 2, 2022, Bend, OR, USA}
\acmPrice{15.00}
\acmDOI{10.1145/3526113.3545627}
\acmISBN{978-1-4503-9320-1/22/10}

\usepackage{balance}    %
\usepackage{color}
\usepackage{colortbl}

\usepackage{multirow}

\usepackage{xspace}

\usepackage{graphicx}
\usepackage{xcolor}

\definecolor{Gray}{gray}{0.97}
\definecolor{MedGray}{gray}{0.9}
\definecolor{greytext}{gray}{0.5}
\definecolor{DarkGreen}{rgb}{0.0, 0.5, 0.0}
\definecolor{PFGreen}{rgb}{0.0, 0.5, 0.0}
\definecolor{lightGreen}{rgb}{0.8, 0.9, 0.8}
\definecolor{CadmiumGreen}{rgb}{0.0, 0.42, 0.24}
\definecolor{DarkKhaki}{rgb}{0.74, 0.72, 0.42}
\definecolor{DarkRed}{rgb}{0.7, 0.2, 0.2}
\definecolor{Purple}{rgb}{0.7,0.0,0.7}
\definecolor{Brown}{rgb}{0.7,0.3,0}
\definecolor{Orange}{rgb}{1, 0.5, 0.1}

\graphicspath{
{figures/} %
}

\usepackage[per-mode = symbol]{siunitx}
\DeclareSIUnit\rpm{rpm}
\DeclareSIUnit\fps{fps}
\usepackage{xcolor}

\begin{document}

\title{TangibleGrid: Tangible Web Layout Design for Blind Users}

\author{Jiasheng Li}
\email{jsli@umd.edu}
\affiliation{%
  \institution{University of Maryland}
  \city{College Park}
  \state{Maryland}
  \country{USA}
  \postcode{20742}
}

\author{Zeyu Yan}
\email{zeyuy@umd.edu}
\affiliation{%
  \institution{University of Maryland}
  \city{College Park}
  \state{Maryland}
  \country{USA}
  \postcode{20742}
}

\author{Ebrima Jarjue}
\email{ebjarjue@terpmail.umd.edu}
\affiliation{%
  \institution{University of Maryland}
  \city{College Park}
  \state{Maryland}
  \country{USA}
  \postcode{20742}
}

\author{Ashrith Shetty}
\email{ashrith@terpmail.umd.edu}
\affiliation{%
  \institution{University of Maryland}
  \city{College Park}
  \state{Maryland}
  \country{USA}
  \postcode{20742}
}

\author{Huaishu Peng}
\email{huaishu@cs.umd.edu}
\affiliation{%
  \institution{University of Maryland}
  \city{College Park}
  \state{Maryland}
  \country{USA}
  \postcode{20742}
}

\renewcommand{\shortauthors}{Li, et al.}

\begin{abstract}
We present TangibleGrid, a novel device that allows blind users to understand and design the layout of a web page with real-time tangible feedback.
We conducted semi-structured interviews and a series of co-design sessions with blind users to elicit insights that guided the design of TangibleGrid. 
Our final prototype contains shape-changing brackets representing the web elements and a baseboard representing the web page canvas. 
Blind users can design a web page layout through creating and editing web elements by snapping or adjusting tangible brackets on top of the baseboard. 
The baseboard senses the brackets' type, size, and location, verbalizes the information, and renders the web page on the client browser. 
Through a formative user study, we found that blind users could understand a web page layout through TangibleGrid. They were also able to design a new web layout from scratch without the help of sighted people.
\end{abstract}

\ccsdesc[500]{Human-centered computing~Accessibility systems and tools}

\keywords{Accessible web design, tactile feedback, tangible user interface, visual impairment, accessibility}

\maketitle

\section{Introduction}
\begin{figure}[ht!]
  \includegraphics[width=0.48\textwidth]{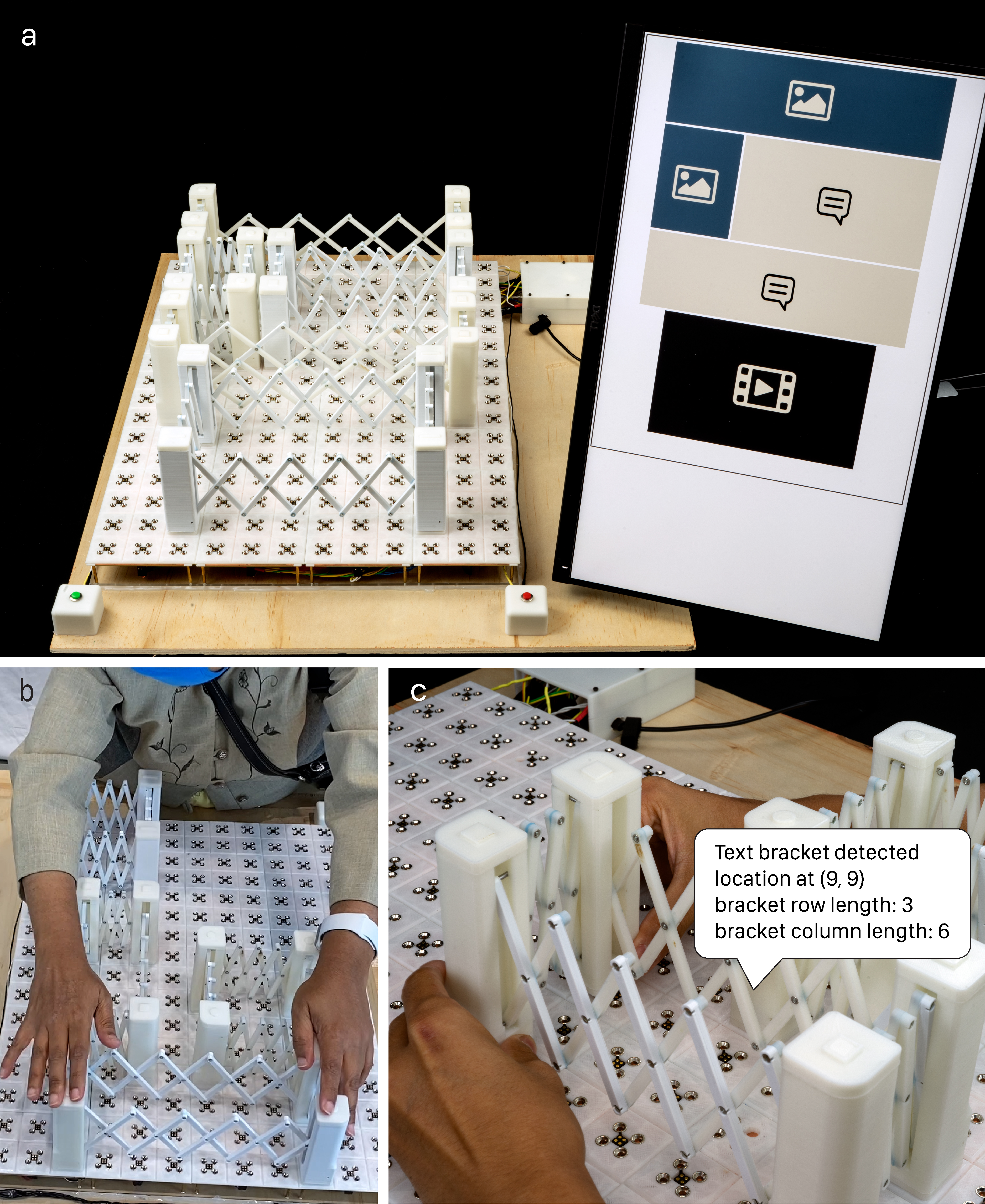}
  \caption{TangibleGrid overview. a) The complete system of Tangible grid; b) a participant is exploring a web page layout; c) designing a new layout by resizing and placing a bracket to the baseboard.}
  \Description{Three photos labeled in a), b), and c) show an overview of the TangibleGrid. Figure a) shows the complete system of the TangibleGrid that includs a baseboard, four brackets on the baseboard, and a PC monitor. Figure b) shows a participant exploring a web page layout with their hands. Figure c) shows a participant designing a new layout by resizing and placing a bracket to the baseboard. Sensing the placement, TangibleGrid provides the audio information of the placed bracket.}
  \label{fig:teaser}
\end{figure}
Assistive technologies have greatly changed the lives of blind and visually impaired people. 
Beyond Internet consumers, blind users are now able to share stories and life events on social media sites such as YouTube~\cite{YouTubers} and Instagram ~\cite{instagram}; some blind users have also created and maintained their own web pages for blogging and knowledge sharing~\cite{dolphin, kearney-volpe_hurst_2021}.
Indeed, the stories and daily experiences of the blind media influencers have become an important source of support to the blind community.
Mastering skills like building web pages has also led to new employment opportunities for blind and visually impaired people~\cite{bell2015employment, Florian}.

Unfortunately, creating a web page is still challenging for many blind users despite the strong need for it~\cite{Editing_Spatial_Layouts_Print, kearney-volpe_hurst_2021}. For one, web page design often requires blind developers to code in HTML and CSS, which has a series of accessibility challenges~\cite{kearney-volpe_hurst_2021}.
Responding to these issues, researchers have proposed workshops and online courses that help blind users learn web programming using screen readers ~\cite{Stefik_2019, kearney-volpe_hurst_fitzgerald_2019}. 
Assistive programming tools such as CodeTalk~\cite{potluri2018codetalk} and StructJumper~\cite{baker2015structjumper} can also help blind users understand the semantic meaning of code structures.
While these efforts support blind users in writing a program or coding web page content, a second barrier is preventing many blind users from having their own web page. Few accessible tools can help blind users understand and design the graphical layout of a web page~\cite{schaadhardt2021understanding}, where visual semantics such as the size, shape, and location of the content matter~\cite{kearney-volpe_hurst_2021, potluri2021examining, Editing_Spatial_Layouts_Print}.

Recently, researchers have started exploring ways of allowing blind users to understand and edit graphical layouts on a screen. 
Potluri \textit{et al.} ~\cite{Edit_Webpage_Designs_Gesture} showcase a prototype that allows blind developers to modify a web page layout by coding in the IDE or using gestures on a touchscreen.
Li \textit{et al.} ~\cite{Editing_Spatial_Layouts_Print} present a multimodal tool that allows blind users to understand a web page layout with tactile print-outs and change it using a self-voicing tablet application. 
While their tool offers tactile feedback for web page layout editing, users must reprint a new layout with swell paper every time a change is made. 
The multiple-step editing process is not as smooth as the direct manipulation approach~\cite{hutchins1985direct} that sighted users experience.

In this paper, we present TangibleGrid, a working prototype that allows blind users to understand and design the layout of a web page with real-time tangible feedback. 
With TangibleGrid, a blind user can place multiple visual elements, such as a textbox, a figure, or a video on a web page canvas by directly snapping the corresponding tangible brackets onto a custom baseboard. (Figure \ref{fig:teaser}a). Each type of bracket has a unique tactile pattern on its top that blind users can understand. The bracket can also be resized while remaining as a rectangle so that a blind user can alter the web page layout by directly resizing or relocating these brackets.
Changes are registered to the baseboard immediately so that the brackets' location, size, and type can be read to the user in real-time. An HTML web page will also be rendered automatically to the user.

TangibleGrid is the first tool that allows blind users to 1) understand the visual layout of a web page and 2) edit the design independently and with instantaneous feedback. The development of TangibleGrid went through an iterative design process. We started by conducting semi-structured interviews with six blind users to understand their challenges when browsing and/or creating web pages, and the potential solutions that have been explored (if any). We then went through three rounds of co-design sessions with a blind developer in our team, to evaluate various physical probes and artifacts, each emphasizing a specific design perspective that may help the layout design and creation.
The final prototype was evaluated in-person with ten blind participants through a formative user study. All blind participants were able to understand the layout of an existing web page through TangibleGrid. They could also create a web page layout with the prototype, despite some having no previous experience in web page design and editing.

In summary, our paper contributes: 1) the investigation of the practices, challenges, and opportunities that blind users have concerning web page layout design and understanding; 2) a working prototype that supports the creation of a web page layout with real-time tangible feedback; 3) a formative user study to evaluate the tool.

\section{Related Work}

Our work builds upon the notions of accessible web programming, interactive tactile graphics, accessible tangible user interfaces, and web layout design tools.

\subsection{Accessible Web Programming}
Several studies ~\cite{mealin2012exploratory, albusays_ludi_2016, Albusays_2017} have uncovered the numerous challenges that blind users face when programming. For example, commercial IDEs such as Visual Studio Code~\cite{visual_studio} and Apple Xcode~\cite{xcode} lack sufficient accessible features; screen readers such as JAWS and NVDA also have compatibility issues with these programming environments, making it difficult for blind programmers to navigate through lines of code. 

To address the accessibility issues, several IDE plugins are developed to support code navigation and debugging~\cite{potluri2018codetalk, baker2015structjumper, stefik_haywood_mansoor_dunda_garcia_2009}. Workshops, courses, and online resources are also developed to help blind developers or students write programs or get familiar with the IDE features~\cite{Kane_2014, Stefik_2019, kearney-volpe_hurst_fitzgerald_2019, kearney-volpe_fleet_ohshiro_arias_hurst_2021, cisco}. 

Although much effort has been made to support accessible programming in general, web programming renders new challenges on top of the accessible programming issue~\cite{kearney-volpe_hurst_2021, potluri2021examining,schaadhardt2021understanding}. As the output of the code, the web page mainly contains visual information; blind developers have no sufficient tools to access the graphic layout of the design and thus, have difficulties in understanding the web page created by themselves. TangibleGrid hopes to address this issue by offering tangible feedback on a web page layout.

\subsection{Interactive Tactile Graphics}
For blind users, tactile graphics are essential to learning and exploreing graphical information such as maps or bar charts. Traditionally, tactile graphics are made with a Braille embosser or printed on swell paper; therefore, the presented information is static and often with a limited amount due to the restrict paper space~\cite{ brock2015interactivity, holloway2018accessible, jacobson1998navigating}. This makes it challenging to provide sufficient information without overly complicating the printed layout~\cite{ tatham1991design}.  
To overcome these limitations, researchers have proposed to offer additional information using sound ~\cite{ baker2014tactile, miele2006talking} and haptic ~\cite{ yu2000haptic}, sometimes referred as interactive tactile graphics~\cite{maucher2001interactive}. For example, Tactile Graphics Helper~\cite{fusco2015tactile}, Talking TMAP ~\cite{miele2006talking}, and Talking Tactile Tablet~\cite{landau2003use} can generate different levels of audio descriptions based on the points of interest that a user touches. These audio annotations allow what could be verbose in printing to be spoken directly to the user. 

In recent years, the democratization of fabrication technology, such as 3D printing, has extended interactive tactile graphics beyond 2D graphics. 
Researchers have used various 3D printed models to teach blind users the concept of visualization~\cite{kane2014tracking}, to allow them to recognize 3D models~\cite{shi2016tickers,shi2017markit}, to create graphic books for blind children~\cite{kim2015toward}, or to make the tactile interfaces of appliances accessible~\cite{guo2017facade}. Most aforementioned printed artifacts have supporting systems or audio tags to speak the information to blind users, but these 3D printed artifacts remain mostly static and less interactive.
In our work, we also utilized 3D printing technology to make the TangibleGrid prototype. Rather than being static, our tangible brackets can be dynamically changed by blind users to meet their design needs.

\subsection{Accessible Tangible User Interfaces}
Interactive tactile graphics provide scaffolding for blind users to understand a wide range of graphical information, but they often lack sufficient features to be responsive in a haptic manner.
Recently, HCI research has explored the use of tangible user interfaces~\cite{Tangible_Bits} to make tactile graphics dynamic and reactive~\cite{mcgookin2010clutching, schneider2000constructive}.
For example, pin-based displays are common approaches to represent information dynamically~\cite{Holybraille, volkel2008tactile, Graphiti}. 
Systems such as HyperBraille~\cite{volkel2008tactile} can render graphical information, such as a web page, onto a matrix of raised pixels.
ShapeCAD~\cite{siu2019shapecad} further extends the concept to support 3D creation. 
Although dynamic and responsive, one common challenge for these pin-based displays is the high cost. A half-page size, pin-based display can cost more than 50,000 USD, which is not affordable to the majority. 

Another type of accessible tangible user interface is based on the metaphor of an active tabletop. 
For example, Tangible Reels~\cite{ducasse2016tangible} combines a tabletop display and a set of retractable tangible reels to allow visually impaired users to construct tangible maps. 
Following step-by-step audio instructions, blind users can replicate a line-dot map with the set of tangibles and then use the creation to understand the specific information related to the reels and nodes. 
Tangible Desktop~\cite{baldwin2017tangible} further explores the concept by replacing the auditory channel with a set of tangible gadgets, which allows novice screen reader users to have a faster task completion time than audio-only systems.
Mobile robots of various forms have also been used as part of the tangible tabletop interfaces to actively guide the user's attention.
For example, Cellulo~\cite{ozgur2017haptic} allows a blind user to hold it in their hands and then actively guides their hand movement for kinesthetic learning or to display autonomous motion.
FluxMarker~\cite{suzuki2017fluxmarker} uses a flat electromagnetic baseboard to mobilize small magnets, which act as dynamic tactile markers to show a blind user certain points of interest.
TangibleGrid takes inspiration from the aforementioned accessible tangible user interfaces. We convert an HTML canvas into a blank tangible baseboard that is similar to the tabletop metaphor. However, our tool's tactile and haptic features, including the shape-changing tangible brackets and the magnetic snapping method, are specifically made to meet the need for a web layout design tool.

\subsection{Web Layout Design Tools for Blind Users}

For sighted people, there is a great amount of research that focuses on web layout recommendations or graphical layout design \cite{DesignScape,Sketchplore,BeyondGrid,dayama2021interactive,todi2018familiarisation}. 
However, the studies that support blind users to understand or design a web page layout are insufficient, with a few exceptions.
Potluri \textit{et al.} ~\cite{Edit_Webpage_Designs_Gesture} develop a prototype that allows blind programmers to edit a web page layout by combining coding with gestures on a touchscreen.
Tactile Sheets ~\cite{avila2018tactile} discusses the concept of overlaying laser-cut paper on top of a touchscreen device to facilitate the understanding of a digital document's layout and logical structure. 
Li \textit{et al.} ~\cite{Editing_Spatial_Layouts_Print} apply the concept to web page layout design with a working prototype. Their system requires a user to put a tactile print-out of a web page template on top of a self-voicing tablet. 
The user can then feel the web page layout and indicate the modifications they hope to make. The updated layout will be printed out on a different piece of swell paper and then overlaid on the touch screen. The user is then able to confirm the design or work on further editing.
One challenge for overlaying printed layouts on a touchscreen device is that the feedback is not synchronous. 
There is a time delay for each design iteration that requires the user to print a new layout and align it to the touchscreen.

In our work, we share the same promise to support web page layout design by blind users. 
Unlike previous work, TangibleGrid allows blind users to understand and design a web layout in real-time; the user will be able to hear the audio description and confirm the design with their hands every time they add or edit an HTML element on the canvas.

\section{Understanding the Challenge}
To understand the current practices and needs for accessible web layout design tools, we conducted semi-structured interviews ~\cite{adams2015conducting} with six blind users, including one co-author of our paper, who had previously studied web programming at the college level, and maintained his own web page. The demographic information of the rest of the participants is presented in Table~\ref{tab:my-table} as P1 - P5. Among all participants for the interviews, three have self-reported web design or programming experience before or after losing sight; three have no relevant experience. We hoped to understand the challenges that people with different web design literacy levels encounter. Each interview lasted 40 min to 1 hour. We present key findings from the interviews, which, together with insights from previous literature, guide the design of TangibleGrid.

\subsection{Findings}

\subsubsection{Audio is primary}
Participants affirmed that screen readers were the primary assistive tools for most digital activities. They used screen readers to consume web content on PC and phones; some participants also reported using screen readers for productive tools such as PowerPoint and WordPress. Participants mentioned that screen readers were helpful in putting text information in slides or a web page template. They were able to identify empty text boxes with voice guidance. However, all participants mentioned that the voice support could only help them understand what was on the screen (\textit{i.e.}, text box), but not where.

\subsubsection{No accessible web page layout information}
Missing support for web layout understanding was one theme that came across all participants with or without relevant experience. Participants with web programming experience struggled with understanding spatial information, even if they generated the content. As described by one participant: \textit{"...I cannot do anything that is graphical ... I can, you know, do my HTML, CSS, and put all the colors the way I learned it. But I don't know what is on the screen".} P3 talked about their experience of using templates on WordPress and Medium.com. \textit{"...I can put the content in them with the help of screen readers, but I don't know how they are presented on a web page, and I always have to ask a sighted friend to confirm the result".} Other participants without web design experience reported their practice of how they consume web information. To them, missing location information of web content was also frustrating. P2 said, \textit{"I know there is an address bar on the top of the screen, maybe, I think. However, I don't know where everything is on a web page. I just hear them as the computer speaks but that doesn't tell me where exactly on the page ... I have no sense of spatial information where the things are".}

\subsubsection{Desire for autonomy}
Three participants discussed their willingness to be independent. One theme that participants repeatedly brought up was the reliance on sighted people to ``confirm the design'' or to help fix the errors when programming  (\textit{e.g.} counting indentations). %

\subsection{Design Implications}
The interview confirmed the lack of support for blind users --- with or without relevant experience, to understand and design the layout of a web page. From our collected semi-structured interviews, we came up with three design considerations.

\begin{enumerate}
\item \textit{Direct representation of the graphical layout}. 
As suggested in the interview, blind users have no direct way of knowing how web elements are graphically presented on the screen. While screen readers can partially read the context, they cannot adequately describe information such as the location, size, or type of web content. 
Inspired by previous work on accessible tangible interfaces~\cite{mcgookin2010clutching, schneider2000constructive, siu2018shapeshift}, we propose to use tangibles to represent critical visual elements of a web page layout. Ideally, the tangibles should be easy to understand and operate and can offer support to blind programmers and novices who share an interest in creating personal web pages.

\item \textit{Supporting layout design with autonomy}. 
As previously noted, blind users prefer to reduce their reliance on sighted helpers when possible. The web page design can also be personal and may require frequent changes. Thus, we hope our tool can enable blind users to generate the page layout individually.

\item \textit{Multimodal feedback}.
As blind users universally rely on voice feedback, it should be combined with haptics to provide a detailed description.

\end{enumerate}

\section{TangibleGrid}

\begin{figure}[hbt!]
  \includegraphics[width=0.45\textwidth]{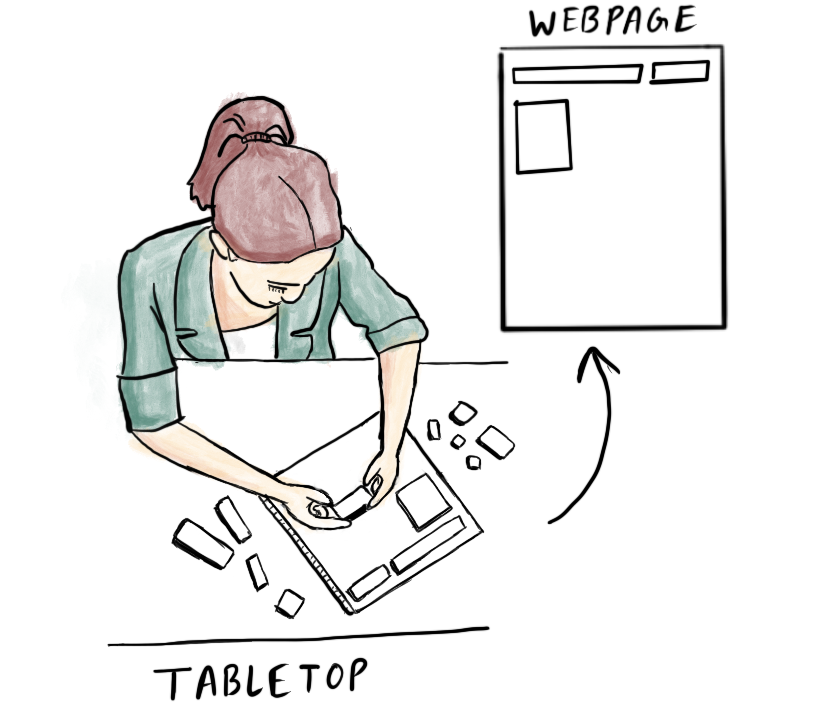}
  \caption{Design concept.}
  \Description{A hand-drawn image shows the design concept. The image shows a person placing physical objects on the baseboard, and a screen shows the web page layout.}
  \label{fig:early_design_concept}
\end{figure}

Informed by the design criteria, we developed TangibleGrid, a novel device that allows blind users to understand and design the layout of a web page with real-time tangible feedback. Figure~\ref{fig:early_design_concept} shows the main design concept of TangibleGrid. The key is to use custom tangibles to represent the graphical layout of a web page. With each tangible representing the main information block of a web page (\textit{e.g.} a text box or a figure), TangibleGrid can allow blind users to understand the overall structure of a web page by scanning across the device with their hands. As these tangibles are also resizable and relocatable, blind users can design the entire web page layout by themselves without constantly seeking help from sighted companions. The audio support of TangibleGrid will verbalize each tangible's location, size, and type, providing blind users real-time feedback on the creation.

Note that since the overarching goal of the tool is to provide a tangible approach to understanding and designing the layout of a web page, we do not intend to physicalize all web page details. In fact, as discussed in ~\cite{edman1992tactile}, simplification is mandatory for tactile exploration. In the case of web layout, we focus on representing the location, size, and types of major web page building blocks. Other information is intentionally omitted in this implementation but can potentially be added as a separate process, which we discuss in Section \ref{advanced_feature}.

\subsection{Design Process}
The design of TangibleGrid is in deep collaboration with Ebrima, the third author of our paper, who is a blind student and researcher in the field of accessibility and, as noted previously, maintains his own web pages. We went through three co-design activities to explore proper tangible mechanisms, tactile patterns, as well as audio feedback that are accessible. Below we briefly describe the three co-design activities and the lessons we learned from them.

\begin{figure}[ht!]
  \includegraphics[width=0.39\textwidth]{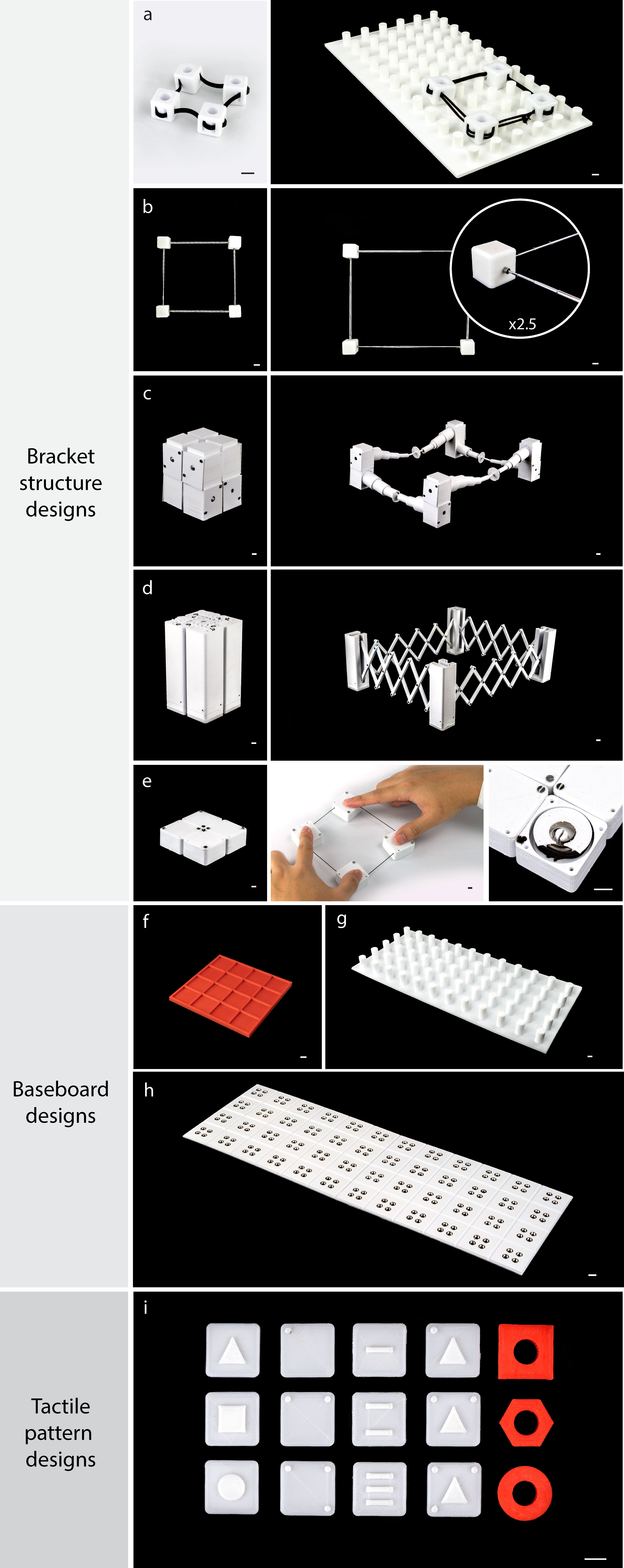}
  \caption{Design probes for the co-design sessions. Figure a) - e) are bracket designs with different connecting methods, including rubber, one-directional telescopic extension, two-directional telescopic extension, scissored linkage, and spring-loaded strings; f) - h) are baseboard designs with raised boarders, extended pillars, and concave grooves with magnets; i) has five different tactile pattern designs, with extruded shapes, Braille-like dots, extruded bars, extruded bars with dots, and side-raised indicators.}
  
  \Description{Eight photos labeled from a) to i) show design probes for the co-design sessions. Image a) shows a tangible bracket with rubber edges in a folded and extended position; image b) shows an extended telescopic bracket in a folded and extended position with a zoom-in position to show the connection place between the telescopic bar and the corner; image c) shows a 3D printed telescopic bracket in a folded and extended position; image d) shows a tangible bracket with scissor structure in a folded and extended position; image e) shows a tangible bracket with spring-loaded string in a folded and extended position with a zoom-in position to show the inside structure of the coil-spring; image f) shows a physical baseboard with the raised grid; image g) shows a physical baseboard with pillars; image h) shows a physical baseboard with concave groove and magnet. Image i) shows five different types of tactile patterns designs. The first left column is different geometrical shapes, triangle, square, and circle. The second left column is one dot, two dots, and three dots. The third left column is one horizontal bar, two horizontal bars, and three horizontal bars. The fourth left column is triangle plus one dot, triangle plus two dots, triangle plus three dots. The fifth column is a square with a hole in the middle, a hexagon with a hole in the middle, and a circle with a middle hole.}
  \label{fig:design_probe}
\end{figure}

\subsubsection{Co-design \#1}
The first round of co-design activity mainly focused on exploring how tangible mechanisms can represent HTML visual elements. As these visual elements are digital bounding boxes of different sizes, our design intuitive was to use resizable rectangle brackets to represent these elements. Specifically, we 3D printed two set of resizeable brackets, using elastic rubber bands and telescopic bars as the resizing mechanism, as in Figure~\ref{fig:design_probe}a and b. The rubber version can resume its initial shape when not used; the telescopic version has rigid linkages to maintain the rectangle shape. Two additional baseboard designs were also prepared to represent the web page canvas as in Figure~\ref{fig:design_probe}f and g, with raised grid edges and extruded pillars as the brackets anchoring mechanisms. 

During the co-design activity, the two bracket probes were presented to Ebrima one by one. 
For each bracket, Ebrima was instructed to first extend or minimize it multiple times, then place it on the baseboards and at different locations. 
Throughout the design session, Ebrima employed the think-aloud method ~\cite{Thinking_aloud, Think-Aloud_Protocols} and talked about if any features of these designs helped or prevented him from 1) understanding the spatial information and 2) moving the bracket from one place to another. 

\noindent\textit{Findings}: Ebrima confirmed that he could understand and change the location and size of the two bracket probes, indicating that representing the web elements with tangible artifacts is a feasible approach. 
Specifically, Ebrima preferred the bracket design with telescopic structure over the rubber band one, citing that the former could provide a rigid feeling and made him feel the edge of the bracket.
Ebrima also commented that putting these brackets onto the two baseboards was challenging since he needed to align all four corners of the bracket to the baseboard pillars or grids for a successful placement. 
However, Ebrima liked the raised grid feature (Figure~\ref{fig:design_probe}f), as these raised edges could allow him to quickly count and find the locations.

The feedback from Ebrima informed the rest of the probes design, which were examined in the second design session.

\subsubsection{Co-design \#2}
In the second round of the design activity, we presented three additional bracket designs: a two-direction telescopic mechanism (Figure~\ref{fig:design_probe}c) as an upgrade to the original telescopic probe, a scissored linkage structure (Figure~\ref{fig:design_probe}d), and a spring-loaded string structure (Figure~\ref{fig:design_probe}e), inspired by Tangible Reels~\cite{ducasse2016tangible}. We also presented a third baseboard design as in Figure~\ref{fig:design_probe}h. The design was inspired by the raised edge feature as in Figure~\ref{fig:design_probe}f, but with engraved grooves and magnets to assist alignment.

Additionally, we prepared five types of tactile patterns as markers to represent different web element types, such as text, figure, and video. The design rationale was to use distinguishable shape features to differentiate web element types. From left to right of the Figure \ref{fig:design_probe}i, the first set of patterns aimed to use extruded shapes as tactile markers; the second and third sets used Braille-inspired dots, and Directional Tactile Paving-inspired bars as tactile markers; the fourth combined set one and two to explore if the combination of different tactile features were recognizable; and the fifth set had raised side edges that explored whether a blind user could recognize the shape by directly grabbing the brackets.

Like before, Ebrima was instructed to test the placement of the bracket probes, feel the tactile patterns, and describe the potential issues of each design. 

\noindent\textit{Findings}: Ebrima confirmed that the scissored linkage bracket (Figure~\ref{fig:design_probe}d) was the best among all bracket candidates, as the scissored linkage was the easiest for him to resize due to its rigidity. The tall height of the design, which we originally thought was a limitation, turned out to be a good feature as it provided a sizeable graspable area for Ebrima to hold the bracket. The new baseboard was also applaudable. \textit{"I like the magnet baseboard. First, it's all flat, feels like an empty HTML file. Second, the magnet force makes bracket snapping feeling good"}. Among the five sets of tactile patterns, Ebrima confirmed that both set 1 and 3 could be recognized easily. The patterns were hard to recognize for set 2 or 4 due to the smaller dot size and the closer distance between patterns. Set 5 was also hard to distinguish. Thus, our final design, as we will introduce in Section~\ref{system overview}, used both set 1 and 3 as the tactile patterns.

\subsubsection{Co-design \#3}
In the last session, we focused on potential voice feedback that can assist the web layout design. A set of audio files were prepared, with three different speed rates, 120, 170, and 220 wpm, as suggested by the previous literature ~\cite{human_listen_rate, ncvs}, and six different content orders, with the type, dimension, and size being first, respectively. Following the Wizard of Oz method~\cite{dahlback1993wizard}, we played these audio files to simulate the auto-generated audio instructions when Embria placed a bracket on the baseboard. We then asked Ebrima to repeat each audio content as accurately as possible. 

\noindent\textit{Findings}: Ebrima could repeat the content at all speeds, with 170 wpm being the most comfortable. 
Regarding the content order, Ebrima commented that the bracket type was most important to him, followed by location and size. Hence the speed and order were chosen for the final design. 
Through the co-design session, we also learned that blind users had to be able to hear a piece of information repeatedly. For example, during the co-design session, we only played the audio description of a bracket placement once. Ebrima pointed out that he might need to hear the description for confirmation repeatedly. He also pointed out that he may hope to know where the previous brackets are when designing a web layout with several brackets. The information on the current bracket solely would not be sufficient. These feedbacks were incorporated into the final system design.

\subsection{System Overview} \label{system overview}
The three rounds of co-design sessions offered us a plethora of insights to effectively translate and tangibly present visual layout, which contributed to our final prototype.
We now detail the main features and the implementation of the tool.

\begin{figure}[htb]
  \includegraphics[width=0.47\textwidth]{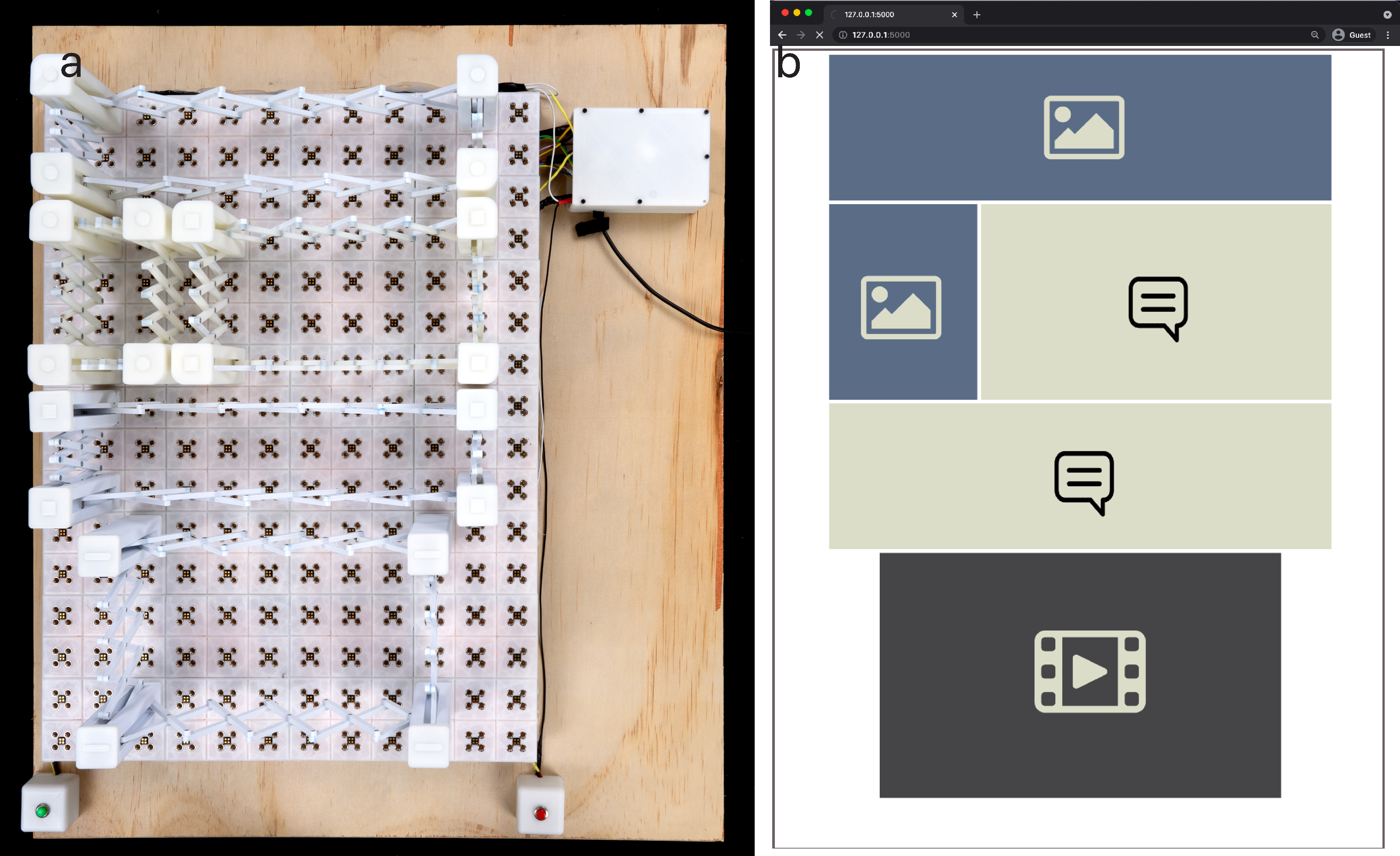}
  \caption{TangibleGrid overview. a) A set of brackets with different types being placed on the baseboard. b) The corresponding web page layout is rendered in a browser.}
  \Description{One photo and a screenshot show the system overview. Image a) shows a set of brackets with different types being placed on the physical baseboard; image b) shows the corresponding web page layout is rendered in a browser.}
  \label{fig:final design}
\end{figure}

Figure ~\ref{fig:final design} shows the TangibleGrid tool.
TangibleGrid composes a physical baseboard and a set of shape-changing tangible brackets representing three essential web page elements: text, figure, and video clip.
The brackets are constrained to the shape of rectangles to reflect the rectangular shape of a web element. 
When a blind user brings a bracket of a particular content type, say an image element, close to the board, it will firmly snap to the baseboard and self-aligned to the grids.
The physical baseboard senses its type, location, and size and immediately speaks this information out to the user. 
The corresponding HTML element is simultaneously rendered on the screen with a content template.
The user can adjust the size or location of the web element by pulling or pushing the corners of the corresponding bracket. The updated information will be vocalized, and the screen will be updated automatically.
If the user hopes to repeat the last bracket's information, they can press the physical button at the bottom right of the baseboard.
The user can press the physical button left of the baseboard to hear the information about all existing brackets.

\begin{figure}[htb!]
  \includegraphics[width=0.47\textwidth]{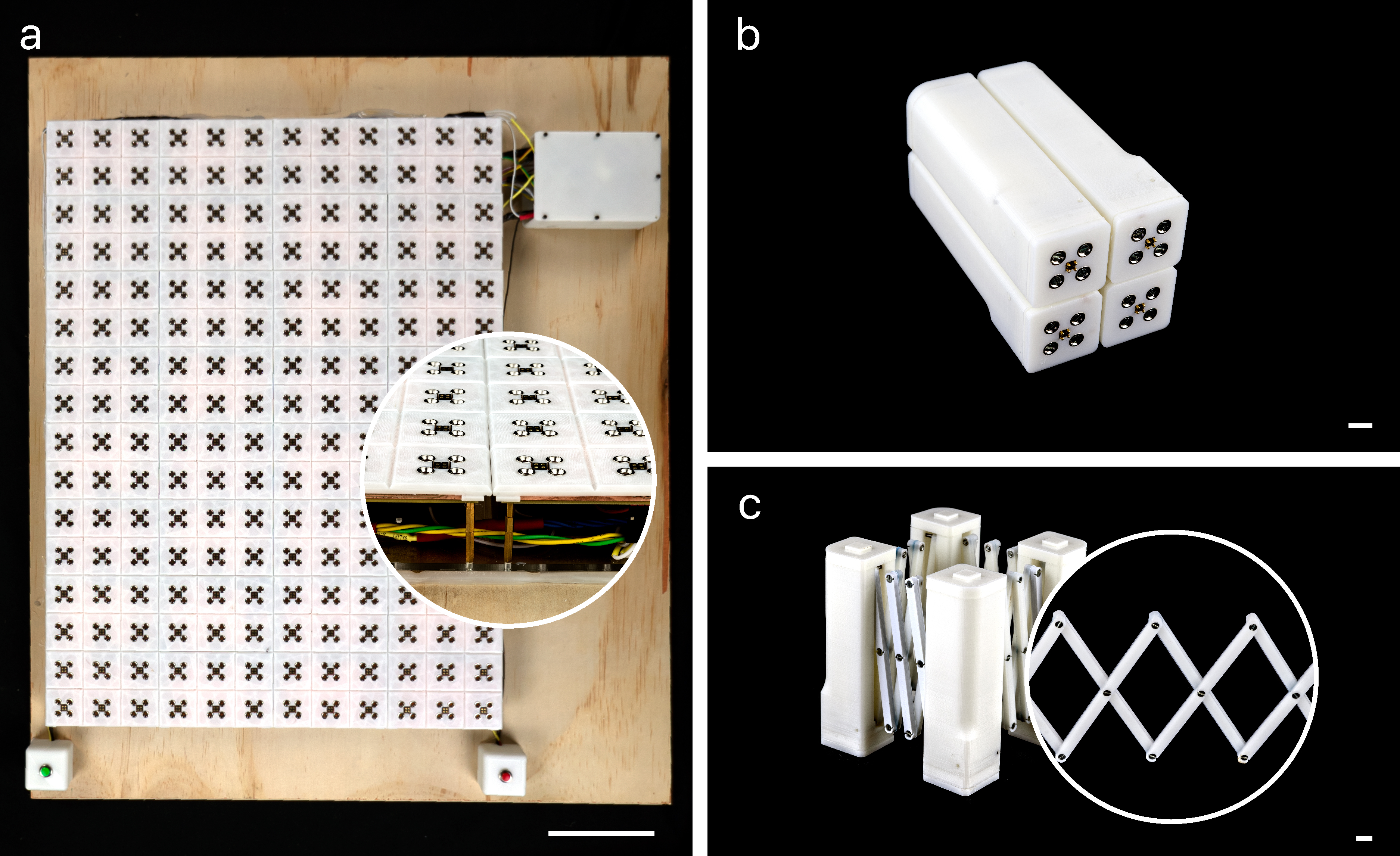}
  \caption{Final prototype. a) The baseboard (scale bar: 50 mm). b) The bracket with magnet base and pogo pin connectors (scale bar: 10 mm). c) The scissored linkage mechanism (scale bar: 10 mm).}
  \Description{Three photos labeled from a) to c) show the final TangibleGrid prototype. Image a) shows the 12 by 16 baseboard with a zoom-in position to show the electronic wiring under the baseboard; image b) shows the bracket with magnetic baseboard and pogo pin connections; image c) shows the scissor structure mechanism.}
  \label{fig:hardware}
\end{figure}

\subsection{Hardware}

As shown in Figure~\ref{fig:hardware}a, the baseboard is \SI{420}{\milli\metre} $\times$ \SI{560}{\milli\metre} with a grid of 12 columns and 16 rows. 
The 12-column design follows the W3C guideline ~\cite{W3C}.
The 16 rows ensure the baseboard has enough space if the user would like to design a vertical web page layout beyond a one-screen asset.

The grid is engraved with `V' shape grooves.
Four \SI{6.35}{\milli\metre} diameter countersunk ring magnets are evenly distributed at the center of each grid cell with a \SI{3}{\milli\metre}  space in between.
At the very center of every four magnets, a 2$\times$2 female pogo pin connector is placed. 
Thus, a total of 192 female pogo pin connectors are placed across the baseboard.
These connectors will connect to the brackets electrically when in contact.

\begin{figure}[htb!]
  \includegraphics[width=0.47\textwidth]{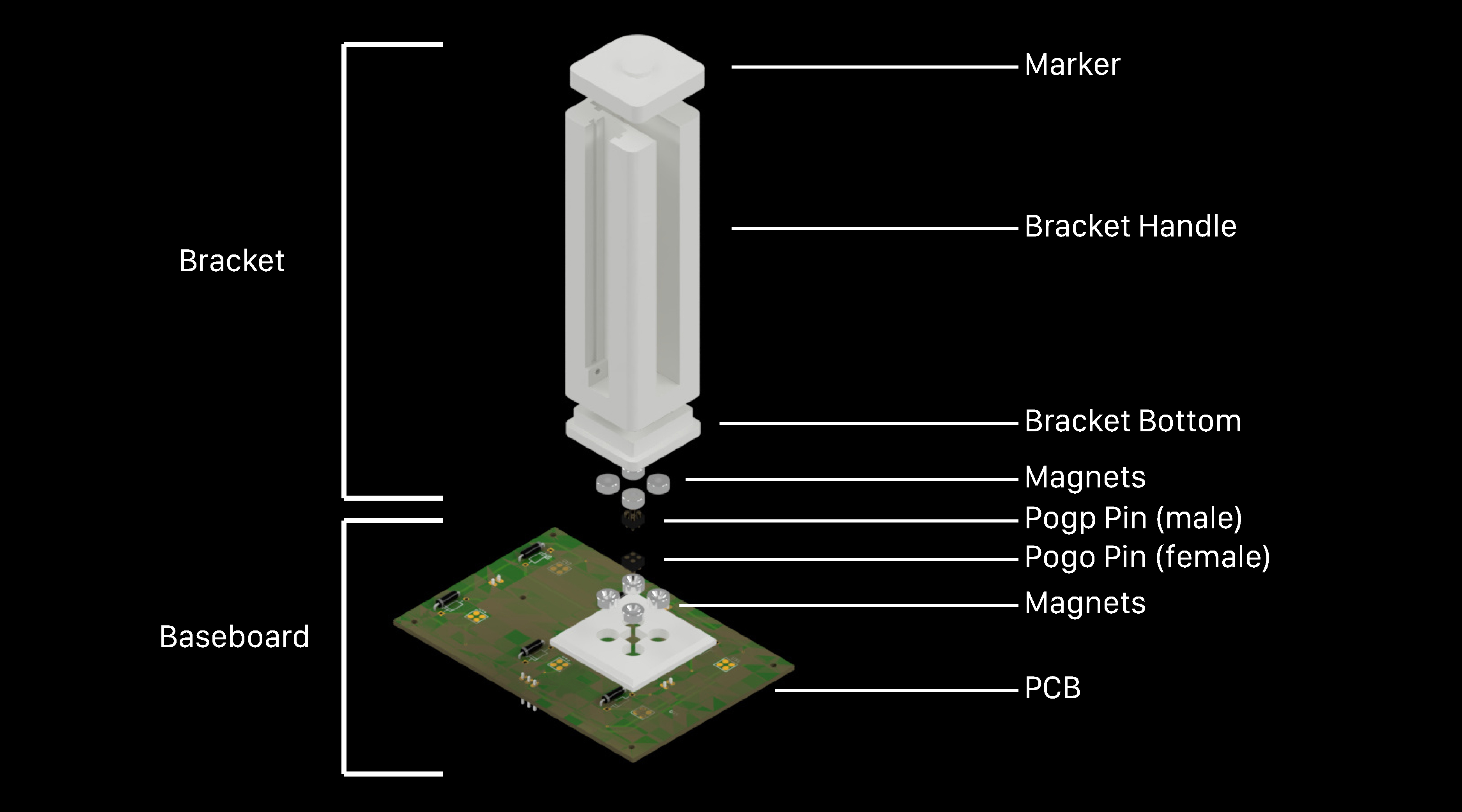}
  \caption{Illustration of the bracket-baseboard connection.}
  \Description{A Illustration to show the bracket-baseboard connection.}
  \label{fig:final base}
\end{figure}

The final bracket (Figure~\ref{fig:hardware}b) has four corner pillars, each with a rounded rectangle square base (r = \SI{3}{\milli\metre}, side length = \SI{30}{\milli\metre}) and a height of \SI{113}{\milli\metre} . 
As suggested by the co-design activities, we employ the scissored linkage to ensure solid, smooth and long-range extension.
Each of the scissored structure has connection links with a size of \SI{91}{\milli\metre} $\times$ \SI{5}{\milli\metre} $\times$ \SI{4}{\milli\metre}. 
They are assembled with M2 bolts and nuts at each revolving joint. 
The bracket has a maximum extension of \SI{420}{\milli\metre} and can be fully folded (Figure~\ref{fig:hardware}c).
Each bracket is equipped with a specific set of tactile markers at its top to indicate the web element's type.
To ensure the bracket type is distinguishable in the software, each corner pillar is inserted with a corresponding resistor to its bottom. The two ends of the resistor are soldered to the two diagonal pins of a 2$\times$2 spring-load male pogo pin connectors that match the ones on the baseboard (Figure ~\ref{fig:final base}). 
Using the 2$\times$2 pin connectors ensures that the resistors in the bracket corners can always connect to the baseboard regardless of the bracket placement orientation.

\subsubsection{Electronics}
\begin{figure}[htb!]
  \includegraphics[width=0.45\textwidth]{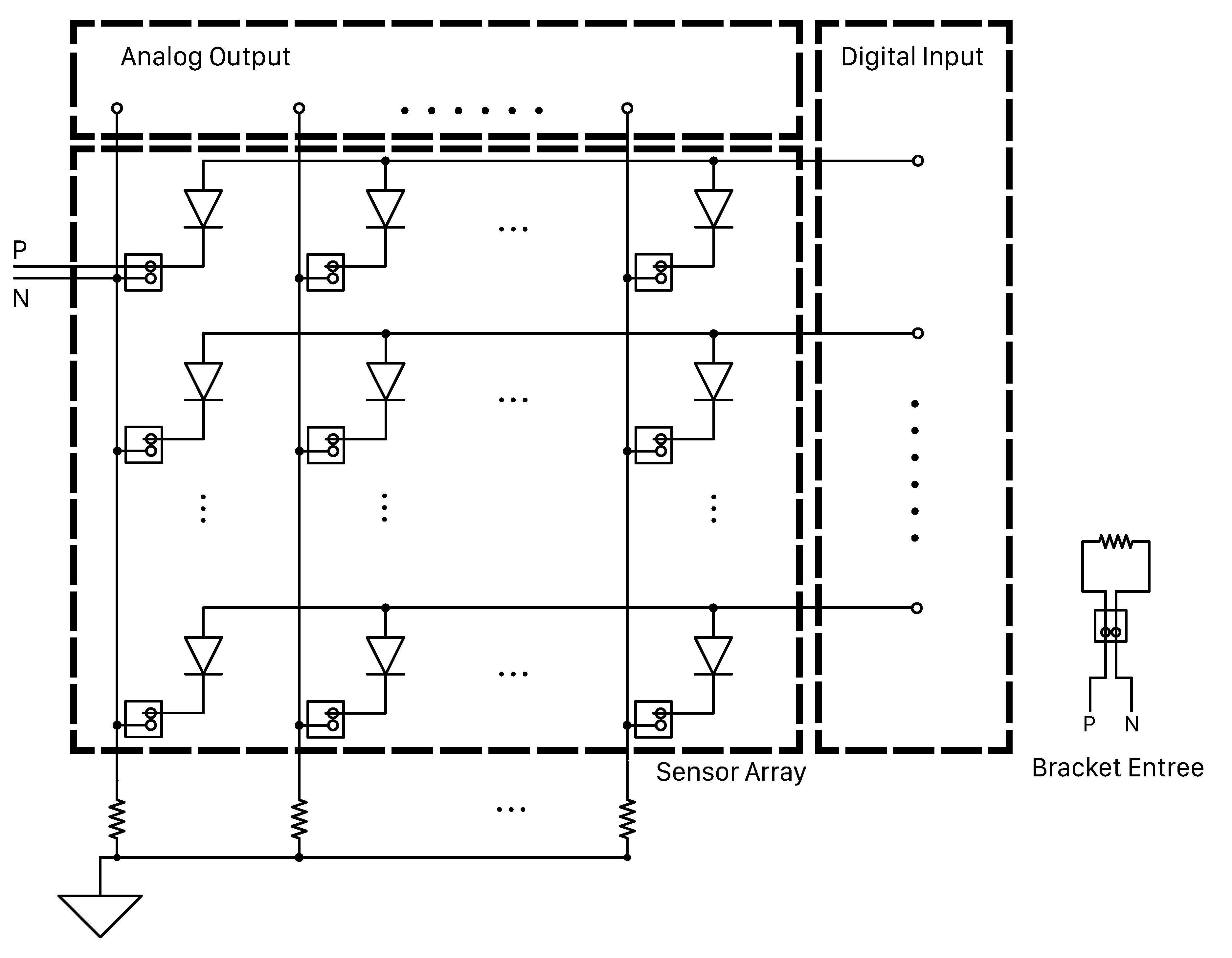}
  \caption{key switch matrix circuit.}
  \Description{A schematic shows the key switch matrix circuit.}
  \label{fig:circuit}
\end{figure}

To recognize the brackets on the baseboard, we implement a key switch matrix circuit (Figure ~\ref{fig:circuit}). 
As each bracket has four corners, and all corners have resistors of the same value, they all act as switches at the intersecting point of the baseboard grid.
When a user places a bracket on the baseboard, the male pogo pin connectors of the bracket is in contact with the female ones on the baseboard, which complete the circuit. 

The scanning algorithm for the key matrix activates one row at a time to detect if any of the column switches are closed.
Multiple closed switches will cause an error known as ghosting or masking, \textit{i.e.}, registering false switch status and failing to detect when a switch isn't closed anymore.
This can be rectified by adding a switching diode in series to the resistor within each switch.
In our case, we use diode 1N4001 to prevent ghosting and faulty readings.

We adopt three resistor values to represent different types of the brackets.
The calculation of the resistance for each placed bracket is performed with a voltage divider circuit. A reference resistor of 1k ohms is used per column to measure resistance values between 180 ohms and 5.5k ohms, with a measurement accuracy of 2.5\%. An Arduino 2560 is used to detect the resistance values of all placed brackets.

\subsection{Software}
Our software application runs on Flask~\cite{flask}, a minimal web framework in Python. The main features of the software are to render the web page layout on the screen and generate audio feedback for the blind user in real time. To render the layout, the software contains a pre-defined web canvas file with the size of 1560 px $\times$ 2080 px. When a new bracket placement is detected by the Arduino, the information is sent to the host software where the web canvas updates the rendering automatically. Meanwhile, the information is also passed to a text-to-speech engine, which then verbalizes the bracket's type, location, and size.

\section{User Study}
We conducted a formative user study to evaluate how the components of TangibleGrid (\textit{e.g.}, tangible brackets, physical baseboard, audio feedback) perform in enabling blind people to understand and design a web page layout.

\begin{table*}[]
\caption{Participants demographics.}
\label{tab:my-table}
\resizebox{\textwidth}{!}{%
\begin{tabular}{@{}ccllllll@{}}
\toprule
\textbf{ID} &
  \textbf{Age} &
  \multicolumn{1}{c}{\textbf{Gender}} &
  \multicolumn{1}{c}{\textbf{Vision Level}} &
  \multicolumn{1}{c}{\textbf{Accessible Aids Software}} &
  \multicolumn{1}{c}{\textbf{\begin{tabular}[c]{@{}c@{}}Programming \\ Skill\end{tabular}}} &
  \multicolumn{1}{c}{\textbf{\begin{tabular}[c]{@{}c@{}}Web Design \\ Experience\end{tabular}}} &
  \multicolumn{1}{c}{\textbf{\begin{tabular}[c]{@{}c@{}}Education \\ Background\end{tabular}}} \\ \midrule
P1  & 56 & Female & Totally blind                     & JAWS                                  & No  & No  & Bachelor             \\
P2  & 46 & Female & Totally blind w/ Light perception & JAWS                                  & No  & No  & Master               \\
P3  & 43 & Female & Totally blind w/ Light perception & JAWS, Braille Display                 & No  & Yes & Master               \\
P4  & 44 & Male   & Legally blind                        & NVDA, Narrator on Windows             & Yes & Yes & Ph.D.                \\
P5  & 43 & Male   & Legally blind                        & Voice-over on iPhone, JAWS on Windows & Yes & Yes & Bachelor             \\
P6  & 32 & Female & Totally blind                     & Voice-over on iPhone, JAWS on Windows & No  & No  & Bachelor             \\
P7  & 67 & Male   & Totally blind w/ Light perception & Voice-over on Mac                     & No  & No  & Bachelor in progress \\
P8  & 50 & Male   & Legally blind                        & JAWS                                  & No  & No  & Master               \\
P9  & 66 & Female & Legally blind                        & JAWS                                  & No  & No  & Master               \\
P10 & 54 & Female & Totally blind                     & Voice-over on iPhone, JAWS on Windows & No  & No  & Bachelor             \\ \bottomrule
\end{tabular}%
}
\end{table*}

\subsection{Participants and Apparatus}
We recruited 10 participants (6 female, 4 male, age 32-67) through online postings (table~\ref{tab:my-table}). 6 participants were self-reported as totally blind; 4 were legally blind. 
All participants were familiar with screen reader technologies such as JAWS or NVDA to help them browse websites daily; one participant mentioned that a Braille display was preferred than a screen reader when browsing websites.
7 participants stated that they did not have any websites design experience; 3 participants stated that they had limited experience either in web page design or in programming.

The study apparatus included one set of TangibleGrid prototype with a 12x16 baseboard, five brackets (two text brackets, two image brackets, one video bracket), and one laptop.

\subsection{Procedure and Tasks}
The user study contained three stages with 90 minutes in total. All participants were compensated at a rate of \$20 per hour in gift card or cash.
In the learning stage, we introduced basic web design concepts as well as the TangibleGrid prototype to the participants. In the following two stages, we asked each participant to complete two tasks: understanding an existing web page layout, and designing a web page layout from scratch. Following each task, we asked participants about their experience by answering Likert scale questions. After participants had completed all tasks, we conducted semi-structured interview to ask them about the overall experience. The study was video and audio recorded for data analysis.

\subsubsection{Learning stage}
After collecting participants' demographics and technology experiences, we presented our prototype to participants. 
Participants were asked to get familiar with the tangibles, for example, by extending or folding the brackets, or by scanning across and touching the baseboard. 
During the learning stage, we explained to the participants how the prototype related to the web page layout, \textit{e.g.}, the baseboard represents a web page canvas and brackets represent web content elements. 
Participants could take time to familiarize themselves with TangibleGrid until they felt comfortable. %
We then started the task 1.

\begin{figure}[htb!]
  \includegraphics[width=0.47\textwidth]{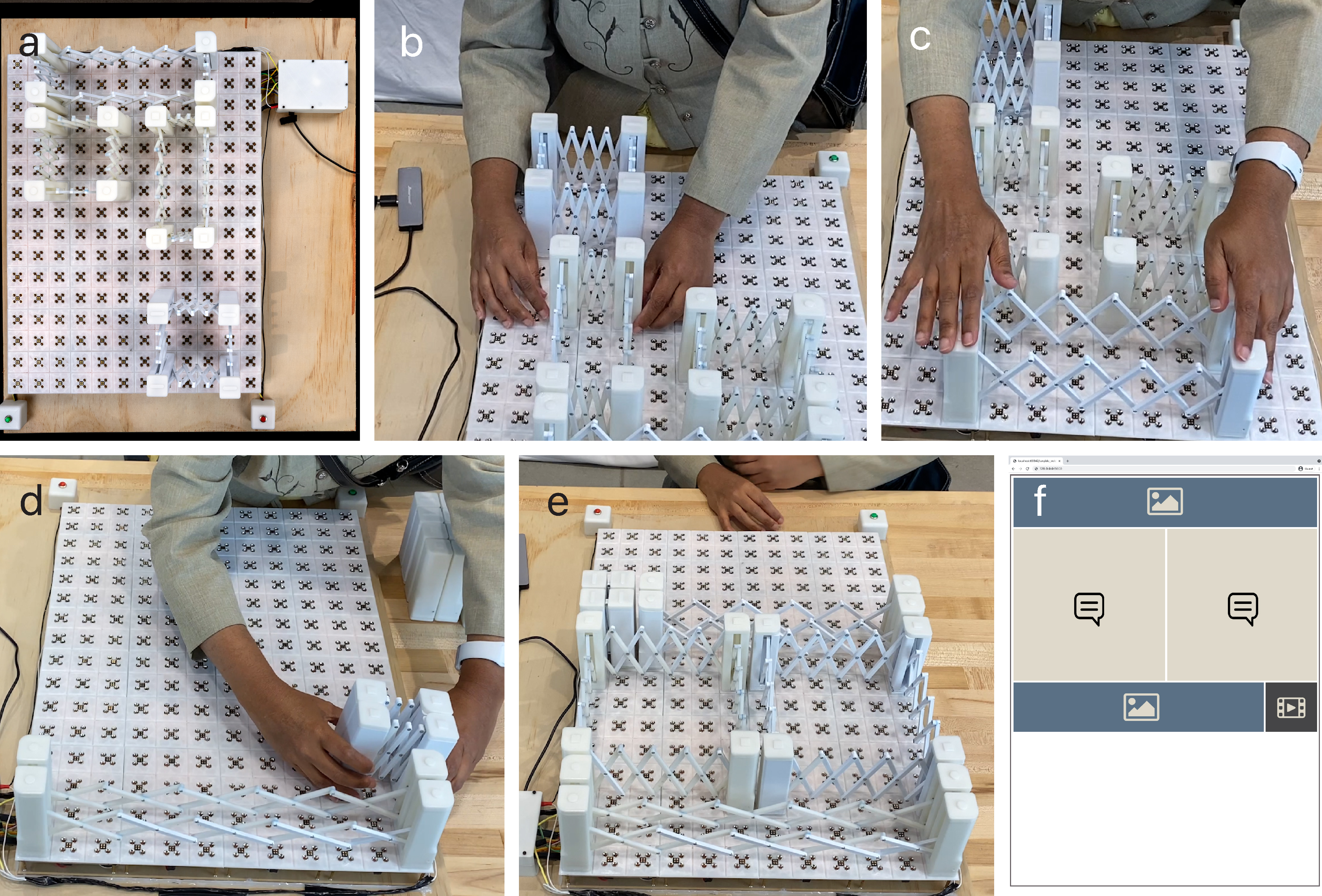}
  \caption{Two user study tasks. a) The web layout template for task 1. b) A participant is counting the size of a bracket. c) A participant is recognizing the type of a bracket. d) A participant is placing a bracket for task 2. e) The completion of the task 2. f) The rendered task 2 web page. }
  \Description{Five photos labeled from a) to e) and a screenshot f) show the two tasks for the study. Image a) shows the web layout for task 1; image b) shows a participant counting the size of a bracket; image c) shows a participant recognizing the type of a bracket; shows d) a participant placing a bracket for Task 2; image e) shows the completion of task 2. The screenshot f) shows the web page layout on the screen.}
  \label{fig:tasks}
\end{figure}

\subsubsection{Task 1: understanding of an existing web layout}
In order to evaluate how the TangibleGrid may help participants understand the web page layout, we presented participants with a pre-defined web page layout template, as is shown in Figure ~\ref{fig:tasks}a. 
The template web page layout contained four brackets with different sizes and types, distributed in a spread-out manner. 
We asked participants to report the corresponding web page layout, including the brackets' size, location, and type.
As we hoped to learn if blind participants could tangibly understand the layout by touching and counting through the brackets and baseboard grid, the audio feedback was turned off for this task. 

\subsubsection{Task 2: web layout design from scratch}
In this task, we investigated how TangibleGrid may allow blind users to build a web page layout by themselves. 
Our original task plan was to have the participants design a web page layout freely. We soon learned from the pilot study that participants’ designs might vary and thus were not comparable. As the goal was not to understand and task blind participants’ design and creativity, we decided to ask all participants to create an identical web page layout. Figure~\ref{fig:tasks}e and f showed the task web page layout. During this task, we told participants the size and location of each web element. Participants had to find the corresponding tangibles and put them on the baseboard by themselves. The audio feature was on during this task.

\subsection{Results}\label{study_result}
We present our user study result in this section and summarize all the participants' findings and feedback. Note that the Likert scale questions are ranged from 1 to 7; 1 refers to strongly disagree, and 7 refers to strongly agree.

\begin{figure}[htb!]
  \includegraphics[width=0.47\textwidth]{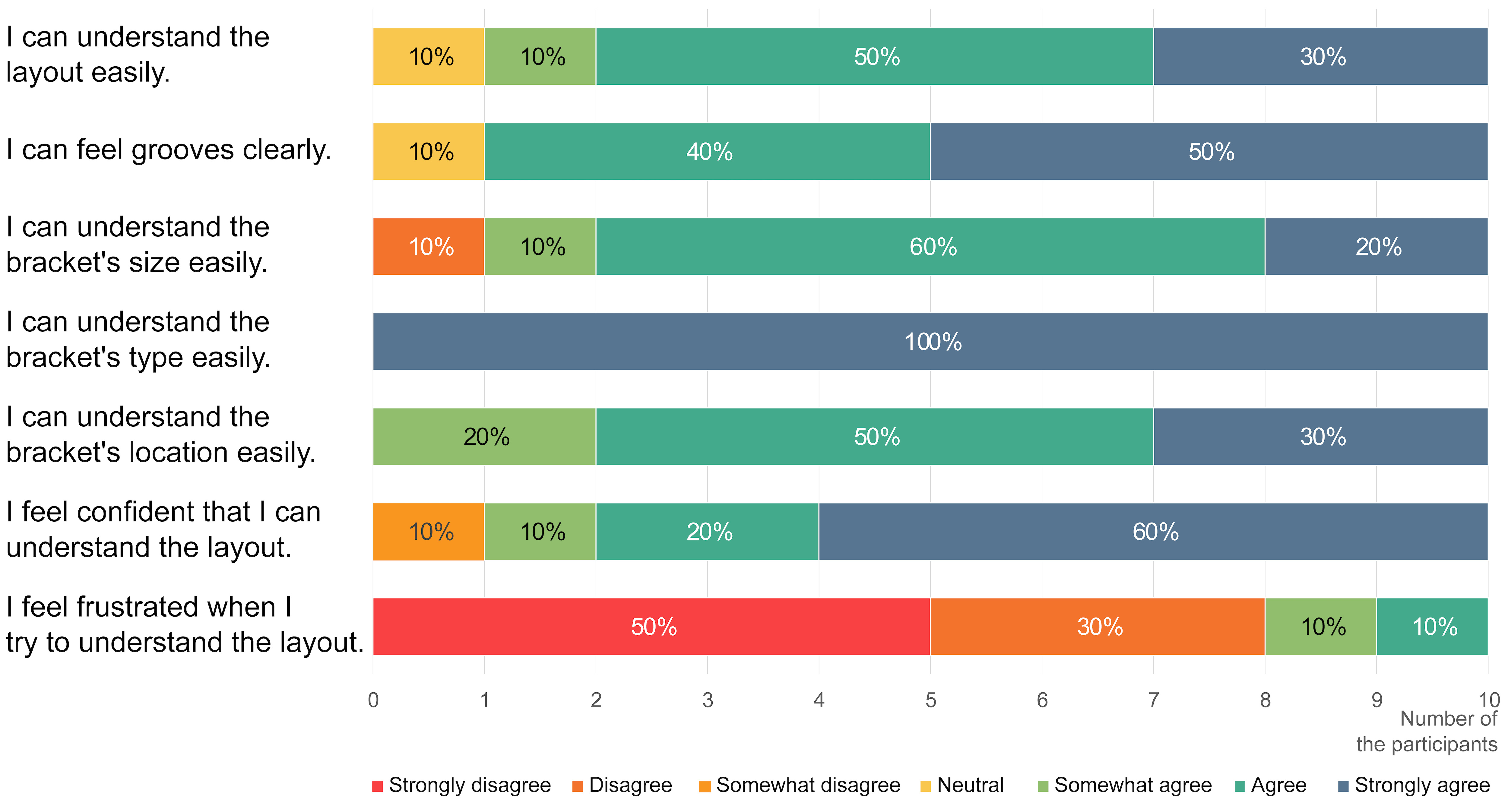}
  \caption{Self-reported ratings of web layout understanding using TangibleGrid.}
  \Description{This image presents a self-reported understanding of web layout using TangibleGrid. Scale from 1-7; 1:Strongly disagree, 2: Disagree, 3: Somewhat disagree, 4: Neutral, 5: Somewhat agree, 6: Agree, 7: Strongly agree. 
The score from question 1, "I can understand the layout easily", three participants report "strongly agree"; five participants report "agree"; one participant reports "somewhat agree"; one participant reports "netural".
The score from question 2, "I can feel grooves clearly", five participants report "strongly agree"; four participants report "agree"; one participant reports "somewhat agree"; one participant reports "netural".
The score from question 3, "I can understand the brackets' size easily", two participants report "strongly agree"; six participants report "agree"; one participant reports "somewhat agree"; one participant reports "disagree".
The score from question 4, "I can understand the brackets' type easily", ten participants report "strongly agree".
The score from question 5, "I can understand the brackets' location easily", three participants report "strongly agree"; five participants report "agree"; two participants report "somewhat agree".
The score from question 6, "I feel confident that I can understand the layout", six participants report "strongly agree"; two participants report "agree"; one participant reports "somewhat agree"; one participant reports "somewhat disagree".
The score from question 7, "I feel frustrated when I try to understand the layout", five participants report "strongly disagree"; three participants report "disagree"; one participant reports "somewhat agree"; one participant reports "agree".}
  \label{fig:result_task1}
\end{figure}

\subsubsection{Task 1: understanding of an existing web layout}
Overall, all the participants were able to correctly report the web page layout during the task. %
As showed in the summarized self-reported rating (Figure~\ref{fig:result_task1}), participants confirmed that they understood the web page layout (\textit{M} = 6.0, \textit{SD} = 0.94) with high confidence (\textit{M} = 6.2, \textit{SD} = 1.3) and low frustration (\textit{M} = 2.2, \textit{SD} = 1.8). 

Key tangible features such as the bracket's type (\textit{M} = 7.0, \textit{SD} = 0), size (\textit{M} = 5.7, \textit{SD} = 1.4), and the grooves (\textit{M} = 6.3, \textit{SD} = 0.95) on the baseboard all contributed to the layout understanding. For example, P5 and P6 pointed out that the grooves were helpful in that they were obvious when scanning with hands. They could quickly sense them and count how many grooves (lines) were in front of a bracket. P9 highlighted that the bracket types and sizes could be understood by touch.
\begin{quote}
\textit{I know the location for each bracket, and how big it is, yeah, I have a sense of them, ..... also, they are image, image, text, and then a video bracket.}  --P9
\end{quote}

One interesting finding during the task was that the scissored structure of the bracket, which was mainly designed to constrain the rectangle shape during resizing, also served as a key tangible scaffold for blind participants.
We observed that on several occasions, participants touched the scissored structure first and then followed through the structure to find four corners of the bracket.
\begin{quote}
\textit{ Recognizing different bracket categories and identifying which four corners belong to one bracket was the first thing I did. This one is easy. I picked any of these (corners), just found one, and traced it (scissored structure) to get to the second corner. ... And I saw these two were not connected, but these two were connected.}  --P2
\end{quote}

\subsubsection{Task 2: web layout design from scratch}\label{result_magnet}
\begin{figure}[htb!]
  \includegraphics[width=0.47\textwidth]{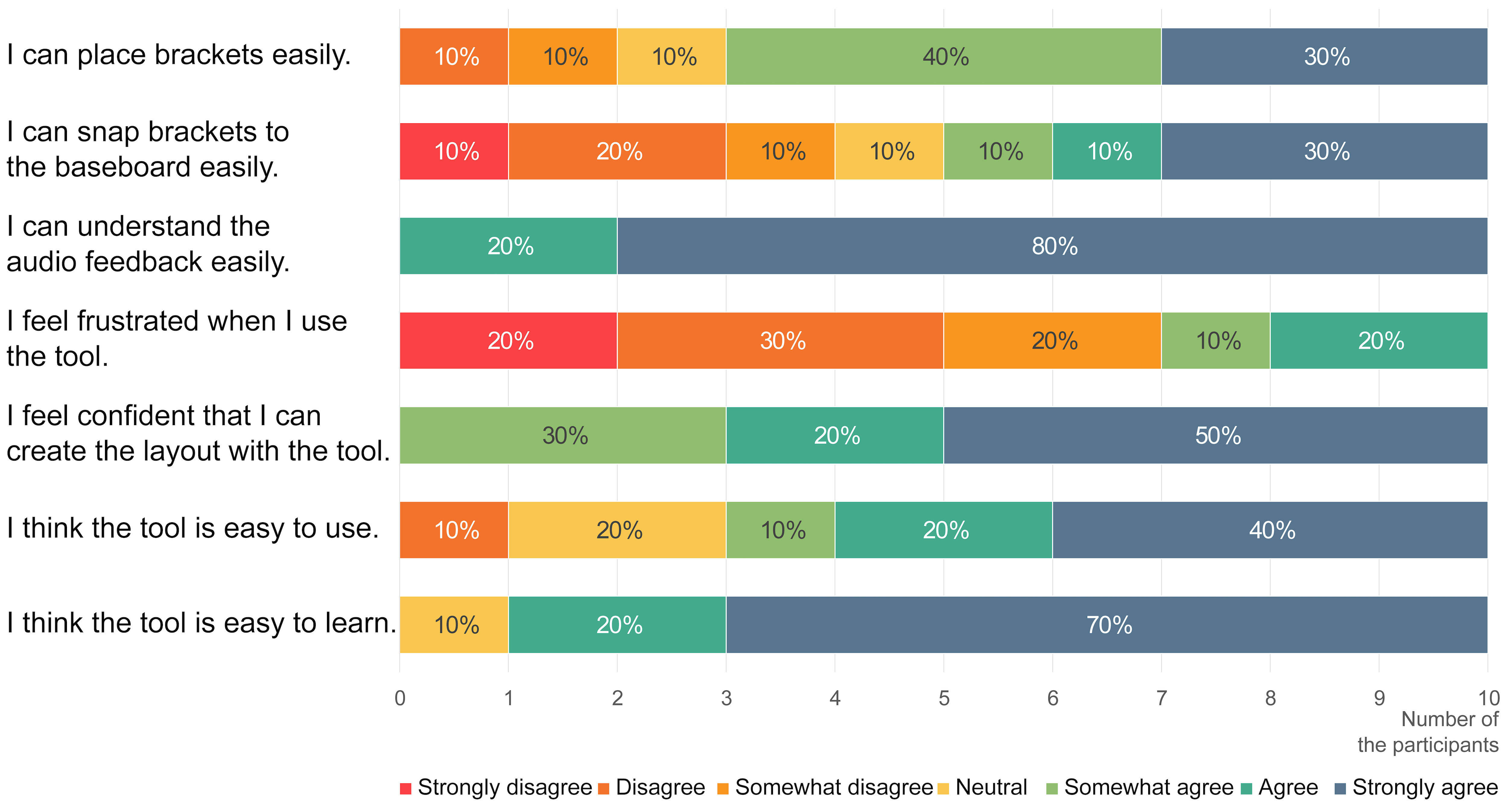}
  \caption{Self-reported ratings of designing a web layout using TangibleGrid.}
  \Description{This image presents a self-reported rating of designing a web layout using TangibleGrid. Scale from 1-7; 1:Strongly disagree, 2: Disagree, 3: Somewhat disagree, 4: Neutral, 5: Somewhat agree, 6: Agree, 7: Strongly agree. 
The score from question 1, "I can placebrackets easily", three participants report "strongly agree"; four participants report "somewhat agree"; one participant reports "netural"; one participant reports "somewhat disagree"; one participant reports "disagree".
The score from question 2, "I can snap brackets to the baseboard easily", three participants report "strongly agree"; one participants reports "agree"; one participant reports "somewhat agree"; one participant reports "netural"; one participant reports "somewhat disagree"; two participants report "disagree"; one participant reports "strongly disagree".
The score from question 3, "I can understand the audio feedback easily", eight participants report "strongly agree"; two participants report "agree".
The score from question 4, "I feel frustrated when I use the tool", two participants report "strongly disagree"; three participants report "disagree"; two participants report "somewhat disagree"; one participant reports "somewhat agree"; two participants report "agree".
The score from question 5, "I feel confident that I can create the layout with the tool", five participants report "strongly agree"; two participants report "agree"; three participants report "somewhat agree".
The score from question 6, "I think the tool is easy to use", four participants report "strongly agree"; two participants report "agree"; one participant reports "somewhat agree"; two participants report "netural"; one participant reports "disagree".
The score from question 7, "I think the tool is easy to learn", seven participants report "strongly agree"; two participants report "agree"; one participant reports "natural ".}
  \label{fig:result_task2}
\end{figure}

All participants were able to complete the web layout design task.
The result was encouraging, given that the participants included ones with web page design experience and also many with no prior knowledge at all.
As showed in Figure ~\ref{fig:result_task2}, participants stated that the prototype was easy to learn (\textit{M} = 6.5, \textit{SD} = 0.97) and use (\textit{M} = 5.5, \textit{SD} = 1.7), and that they felt confident when creating a web page layout (\textit{M} = 6.2, \textit{SD} = 0.92).

After the task, several participants talked about the specific layout they created:
\begin{quote}
\textit{ ...that's a big heading, then some text saying where we are. And then another text and other (image) type. And then the video just a little player.}  --P5
\end{quote}

\begin{quote}
\textit{ I think I have a kind of (layout) in my brain about how it looks like. All the way down there was the video, you know, the banner across the top, the two areas (rows), the texts that are six by six, and then the image that's ten by two, and then the video that's two by two.}  --P9
\end{quote}

\begin{figure}[htb]
\centering 
  \includegraphics[width=\columnwidth]{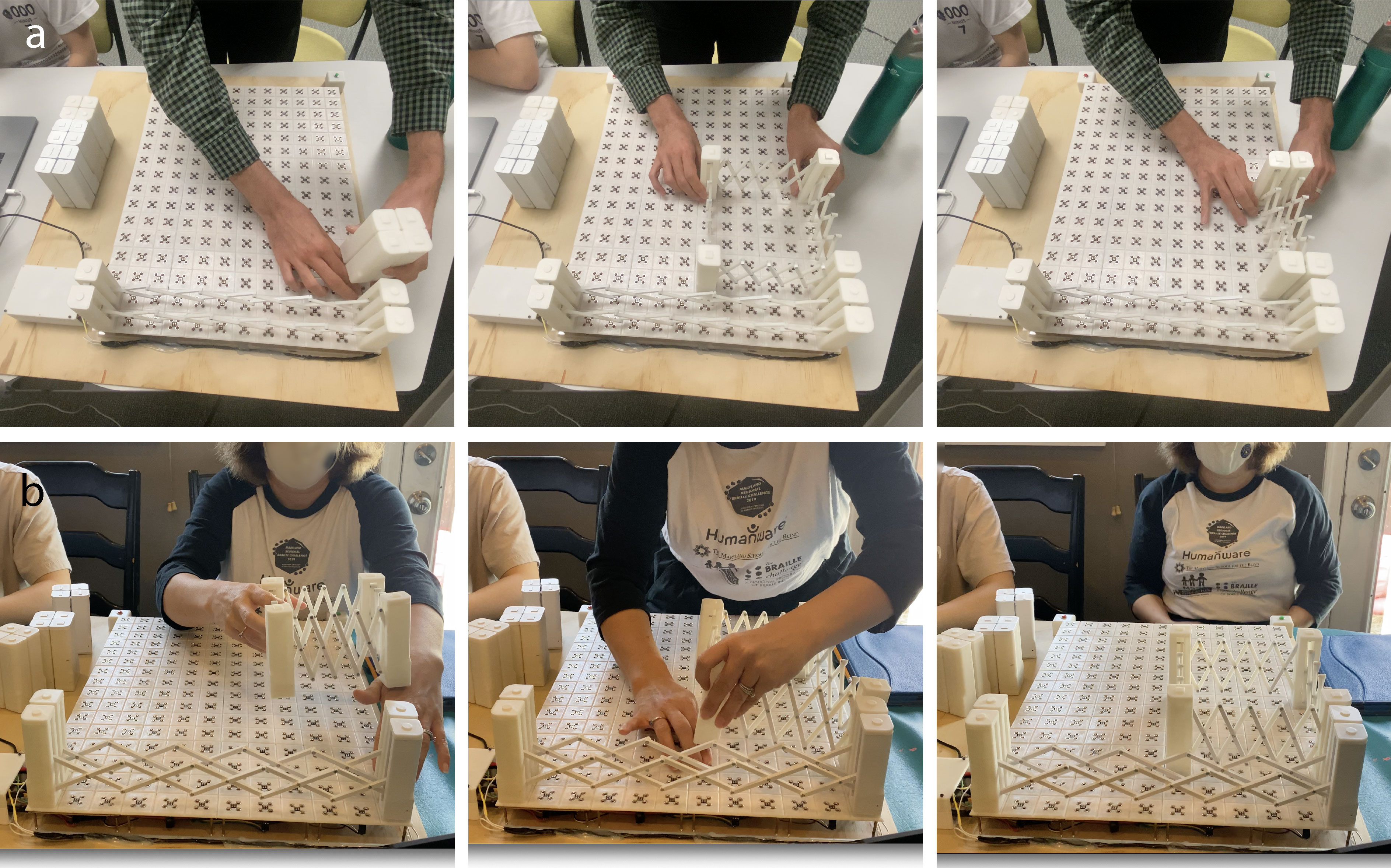}
\caption{Two main strategies that participants used when placing the brackets on the baseboard.}
\Description{Two sets of photos labeled a) and b) show two main strategies that participants used when placing the brackets on the baseboard.}
\label{fig:placement_strategy}
\end{figure}

In terms of the layout building process, we observed that participants had two main strategies when placing the brackets on the baseboard. 
Some participants preferred to place down one corner of the bracket on the baseboard first and then adjusted the bracket's size (Figure~\ref{fig:placement_strategy}a); others preferred to adjust the brackets' size off the baseboard first, and then placed the entire bracket on the baseboard all at once (Figure~\ref{fig:placement_strategy}b).
It is interesting to note that the two different strategies lead to exactly the opposite opinions on the magnet snapping feature (\textit{M} = 4.4, \textit{SD} = 2.3).  
Participants who preferred the first strategy commented the usefulness of the magnet snapping feature, as it helped to hold one corner of the bracket on the baseboard, and thus they could adjust the bracket size without changing its location.

\begin{quote}
\textit{..the snapping is good. Because it is magnetic, you know, it gets there very fast.}  --P1
\end{quote}

\begin{quote}
\textit{ The magnet is strong, but I think it is good, and the magnet helps me...Yeah, I think it would be very difficult without it (magnet) to do it. Magnet helps keep it in place while I am trying to adjust other pieces (corners) of it (bracket).}  --P5
\end{quote}

\begin{quote}
\textit{I think the magnet helps me, it will hold it (the corner of the bracket) in place, so I don't bump it. it's (magnet snapping) probably the best way to do that.}  --P8
\end{quote}

\begin{quote}
\textit{I think I like magnets, yeah, I can see how it works, and maybe more efficient; it's really easy to tell when you have it just right, because they don't move easily. ... I don't want them to weak, so they may just knock over.}  --P9
\end{quote}

For some other participants who preferred the second strategy, the snapping feature might not be very helpful. It could be too strong for them to adjust the bracket size freely, especially when the four corners of a bracket could not be snapped to the correct location all at once. 
For example, P4, who extended brackets first and then placed it on the baseboard, stated that the magnet force made him think consciously about it, and he could not adjust the size of the bracket after he placed it on the baseboard.
\begin{quote}
\textit{ I was confused. How far should I keep it (magnet) from the board to avoid the magnets? If I would hold it up too high then I couldn't count out, but if I would hold it down, then it was all the way around. And then I have to use exert force to detach the magnets again.}  --P4
\end{quote}

The somewhat divided frustration rating (\textit{M} = 3.1, \textit{SD} = 1.9) could also reflect these two opposite opinions. We found the frustration rating was highly relevant to the users' perception of the magnet snapping method. Participants (\textit{e.g.}, P1, P5, P7, P8) who found the magnets useful (strongly agree/agree on "I can snap brackets to the baseboard easily") rated low in frustration (strongly disagree/disagree on "I feel frustrated when I use the tool"). Participants (\textit{e.g.}, P3, P4, P10) who commented that the magnets were too strong reported high frustration. 

When placing brackets onto the baseboard, mistakes happen inevitably. 
During the task, we observed that blind participants were able to correct mistakes by following the grooves on the baseboard. 
As one example, a bracket corner could sometimes be misaligned, where it was placed on the baseboard one row above or below where it should be. 
This would make the TangibleGrid system not recognizing the bracket.
Following the baseboard grooves and the bracket linkage, participants were able to find that these two lines were not in parallel. They could then correct the mistake by re-positioning the bracket corner to the right place.

Finally,  audio feedback of our system was sufficient and also effective (Ave = 6.8, SD = 0.42). It could help participants confirm the bracket placement, reducing their workload. Most participants stated that they could understand the audio content clearly. However, the speed of the audio feedback could be too fast for P9.

\section{Discussion and Future Work}

\subsection{Customization}\label{customization}
While the result of the user study confirmed that, despite their prior knowledge,  TangibleGrid could effectively support blind users to understand and design the layout of a web page, it is crucial to consider the individual differences and preferences among users and build features with greater flexibility.

For example, as discussed in Section ~\ref{result_magnet}, participants may develop their own strategies of placing the brackets. The magnetic snapping feature, which was designed to securely lock the brackets to the baseboard, could also be a source of distraction to some users. 
One potential improvement we can make is to allow participants to decide on the strength of the snapping feature. 
This can be achieved by replacing the current permanent magnets of the baseboard with electromagnets, where the magnetic force can be adjusted.

Similarly, the audio feedback in the current system was limited in that the speed and the information fidelity were pre-defined. Like screen readers, we hope customization can be added to the future software so that participants can decide on their preferred voice-feedback profiles.

\subsection{Advanced Layout Design}
TangibleGrid set a foundation to support basic web layout design in a tangible manner. Moving forward, we expect future research can expand its functionalities. 

For example, Ebrima has tried a smaller size of brackets (15mm X 15mm) in an early exploration, which indicates that the baseboard can have a higher granularity of the grids to support finer brackets placement and adjustment. 

The baseboard design can also be improved with higher flexibility. The fixed number of rows may limit a user's creativity, \textit{e.g.}, if a user hopes to build a long scroll page across multiple screen assets. One possible solution is to modularize the baseboard design, where multiple baseboards can be daisy-chained together. The users will be free from the baseboard size limitation and simply add more baseboards when the design space runs out. 

Finally, we expect the tangible approach may support modern web design features, such as responsive web page layout. For example, by adopting the electromagnets as discussed in Section~\ref{customization}, it is possible to design an active tangible baseboard where brackets can be relocated automatically, similar to FluxMarker ~\cite{suzuki2017fluxmarker}. Such a system will also have a tighter integration to its digital web page representation, as layout changes in the software can be directly reflected with the active baseboard. Of course, how to control the motion of the automated tangibles reliably needs further investigation.

\subsection{Supporting Content Design and Editing}\label{advanced_feature}
During our exit interview, several participants expressed the desire to create web content with TangibleGrid. For example, P7 said, \textit{``I'm definitely interested in consuming more of this. Eventually, I want to put some music (on the web page)...''}. P2 also added, \textit{``...it would be nicer to add color (to it) ... that would make it interesting''}.

Indeed, while the current focus of TangibleGrid is on web layout design, it is only one part of the web design challenges. We consider two possible future directions where content input and editing can also be combined with TangibleGrid.

First, content input and editing can be directly integrated to TangibleGrid as core features. For example, once the user places a text bracket on the baseboard, they can put information directly to it, by long-pressing the top of the bracket and speak to a microphone. Using speech recognition, the voice can be converted to the written content. An image can be inserted in a similar way by matching the user's description to an image from a search engine. For this approach to work, the future tangible brackets need to have touch sensing capabilities on its top. An integrated microphone should also be placed to the baseboard. 

The second approach is to combine TangibleGrid with existing web design platforms or programming IDEs, such as WordPress and Pycharm. For example, it is possible to develop a WordPress plugin, where the layout generated from TangibleGrid can be directly exported to the platform. From this layout, screen readers can recognize the auto-generated space holders. WordPress users can then input the web page content and change their properties such as color and font types. Experienced users can directly program the properties of the web elements with HTML and CSS, with the layout taken care of by TangibleGrid.

\subsection{Beyond Web Page Layout Design}
As discussed in recent work such as ~\cite{schaadhardt2021understanding, potluri2021examining}, visual semantics can be critical for collaboration, navigation, and design. Yet, they are inaccessible to blind users in many scenarios and applications. While TangibelGrid focuses on the layout exploration of a web page, it can potentially be extended for other graphical based tools, such as Microsoft PowerPoint, Apple Keynote, and Google Slides. For example, when the baseboard is placed in a landscape manner, it is possible to simulate the presentation slides, with the text and image brackets used to align the digital content graphically. Note that for these tools, the digital canvas is usually rendered free-form. Thus, the grid-based mechanism may limit the resolution of the design. Whether and how TangibleGrid can be extended to support grid-free creation remains an open question for future research.

\section{Conclusion}
In this paper, we described the design and development of TangibleGrid, a novel tangible device that allows blind users to understand and design a web page layout independently. 
Our design was informed by an initial interview with six blind participants and three rounds of co-design sessions that involved multiple iterations of tangible probes and prototypes.
Our final system used a magnetic baseboard to represent an HTML canvas and a set of shape-changing brackets to represent three types of web elements.
Placing these tangible brackets on the baseboard would activate an audio description of their information, and create the corresponding web page.
In the user study, all participants could use TangibleGrid to understand an existing web page layout, and design one from scratch.
We hope TangibleGrid can enable blind users to share their creativity in the future.

\begin{acks}
This work was supported in part by the New Direction Fund from the Maryland Catalyst Fund. We thank Anup Sathya and Biswaksen Patnaik for their feedback. We thank the anonymous reviewers for their valuable comments.
\end{acks}

\bibliographystyle{ACM-Reference-Format}
\bibliography{sample-base}

\appendix

\end{document}